# Evaluating the Effectiveness of Common Trading Models

Joseph Attia

Brooklyn Technical High School

June 2019

Abstract

How effective are the most common trading models? The answer may help investors realize upsides to using each model, act as a segue for investors into more complex financial analysis and machine learning, and to increase financial literacy amongst students. Creating original versions of popular models, like linear regression, K-Nearest Neighbor, and moving average crossovers, we can test how each model performs on the most popular stocks and largest indexes. With the results for each, we can compare the models, and understand which model reliably increases performance. The trials showed that while all three models reduced losses on stocks with strong overall downward trends, the two machine learning models did not work as well to increase profits. Moving averages crossovers outperformed a continuous investment every time, although did result in a more volatile investment as well. Furthermore, once finished creating the program that implements moving average crossover, what are the optimal periods to use? A massive test consisting of 169,880 trials, showed the best periods to use to increase investment performance (5,10) and to decrease volatility (33,44). In addition, the data showed numerous trends such as a smaller short SMA period is accompanied by higher performance. Plotting volatility against performance shows that the "high risk, high reward" saying holds true and shows that for investments, as the volatility increases so does its performance.

**Table of Contents**

## 1.  Research Question

How effective are the most common technical trading models (moving average crossover, regression, K-nearest neighbor algorithm), and how do they compare against each other? In addition, what are the two most optimal periods for moving average crossover trading?

## 2.  Introduction

On May 1894, a boy, Benjamin Graham, was born to a Jewish family in London. At the age of 1, he moved to NYC, where shortly after his family was plagued with loss and poverty. Nonetheless, he graduated second in his class at Columbia University, and was offered a position to teach, but refused it since he needed a larger income to support his family. So, he went to Wall Street. Benjamin Graham went on to be known as the "father of value investing", and trained many notable disciples including Warren Buffet, one of the most infamous investors of the 21$^{st}$ century.

> "The investor's chief problem – and his worst enemy – is likely to be himself.
>
> -Benjamin Graham

The quote above by Benjamin Graham refers to the constant struggle that emotion and greed play in an investor's strategy. When is enough profit enough? And when does a company we personally like force us to invest even though all the signs say not to invest? This is natural but there have been efforts to avoid losing money due to emotional trading. That is the goal of my paper, to evaluate how effective those methods are, and how to make some of them more effective.

I chose this topic since I developed an interest in the stock market from a young age. In 6th grade, I got together with a few of my friends and a teacher entered us into the Stock Market Game, a program by the SIGMA Foundation with the intent of "[connecting] students to the global economy with virtual investing and real-world learning." For the first few weeks of the game, we were doing very well, placing in the top 10 teams in the country. Out of nowhere, a week before the contest finished the stocks in our portfolio tanked and our position fell alongside them. I returned to the Stock Market Game in the 9th grade as part of my schools Stock Market Club, and this time, with a different team, we peaked at first place a couple of times but ultimately ended in 5th place nationally. The stock market was exhilarating and fascinating, but more importantly, unpredictable.

In 11th grade, I began a discrete mathematics course which began with graph theory. After writing a paper on the Applications of Graph Theory to Investing, I had ended with an attempt to use directed graphs and correlated stocks as an indication of when to invest, but the method provided varied results. When the time came to research again, I didn't hesitate to continue researching trading methods and decided to recreate and explore common technical trading models.

# 3. History

## 3.1. History of Mathematical Models

The use of mathematical models, while not as old as graph theory, have been around for a while. The first distinguishable mathematical models were in fact numbers, which can be traced back to 30,000 BC. Following numbers, astronomers and ancient architects began to apply mathematical models. Around the 20$^{th}$ century, the first computer was released and with it a plethora of ways to model math but even better, a way for everyone to develop countless mathematical models themselves.

### 3.1.1. Use of Mathematical Models in Finance

As the personal computer became popular, people began introducing mathematical modeling to finance. One of the best-known mathematical models applied differential equations and Brownian motion, described by Louis Bachelier in 1900 and Albert Einstein in 1905, in order to estimate the price of a European option. This model is called the Black-Scholes model, was developed in 1973, and was heavily used throughout the following decades.

## 3.2. History of the Stock Market

The first stock exchange was established in Belgium in 1531 where brokers and moneylenders met in order to create deals with businesses, governments, and even individual debt. In the 17$^{th}$ century, charters were granted to companies in many different European countries which granted governments a stake in the profits in the East. Nevertheless, many ship owners began seeking more investors so that if anything happened to their ships while at sea their fortune would not be ruined. Investors would also manage their risk by investing in many different ventures, ensuring profits would cover their losses.

The London Stock Exchange was opened in 1773 and the New York Stock Exchange (NYSE) was opened 19 years after. Unlike the London Stock Exchange, whch was restricted from selling shared, the NYSE sold stocks from its birth. The NYSE was located in one of the most economically thriving cities in America on Wall Street, it quickly became the most popular stock exchange in the country.

In 1971, nearly 200 years after the inception of the NYSE, the Nasdaq was created. Developed by the Financial Industry Regulatory Authority, the Nasdaq was the first of its kind in the world since it didn't take up a physical building. It was a network of computers that executed all trades electronically, which in turn made trading more efficient. This competition forced the NYSE to step up its game by merging with a European stock exchange and becoming the first worldwide exchange.

### 3.3.    History of Technical Analysis

Technical analysis first appeared with the introduction of the Dow Jones Index, as Charles Dow recorded data like the highs and lows of his daily averages, and the weekly and monthly correlating patterns. Charles would try to find patterns leading up to past events and try to predict future events using those patterns. Then another investor, William P. Hamilton, began to take short term waves to filter out the insignificant day-to-day fluctuations and used the direction of averages to confirm the direction of the investment. Using his methods, Hamilton predicted the 1929 crash before it happened. There have been numerous developments to the field in the years after Dow and Hamilton, and its creation cannot be credited to one person, but rather the consistent addition of material to the field.

### 3.4. History of Machine Learning Models

Machine Learning has become a huge part of business and research even though it does not date back more than 70 years. Machine Learning models are based on the brain, a collection cells repeatedly assisting in firing another, but as an artificial neural network with artificial neurons and the effects one has on another. The phrase "Machine Learning" was coined by Arthur Samuel, an IBM developer that created a computer program to play checkers that used many mechanisms including rote memorization with which the computer recorded all previously seen positions and their outcomes. The first neural network, the Perceptron model made by Frank Rosenblatt, combined Samuel's efforts with the brain's cell interaction. In 1967, the nearest neighbor algorithm was created and created a basis for using machine learning for pattern recognition. Since then, many new algorithms have been developed and research about multilayering algorithms and backpropagation has become more popular.

## 4. Mathematical Background

### 4.1. Intro to Investing and Stocks

When analyzing historical stock data, analysts have a variety of different options for which data they would like to use. Ways to represent stock prices include open price (price at beginning of the trading day), close price (the price before the trading day ends), high (the highest point of the stock during the trading day), low (the lowest point of the stock during the trading day), change (the change in stock price, which can be in dollars or percent), volume (the total amount of shares traded for the trading day), adjusted closing price (a closing price that has been adjusted to account for any splits and dividends).

| Date | Open | High | Low | Close | Adj Close | Volume |
| --- | --- | --- | --- | --- | --- | --- |
| 2017/12/1 | null | null | null | null | null | null |
| 2018/1/1 | 1187.32 | 1273.99 | 1187.32 | 1251.42 | 1251.420044 | 73366640000 |
| 2018/2/1 | 1248.27 | 1258.88 | 1120.08 | 1201.87 | 1201.869995 | 79579410000 |
| 2018/3/1 | 1202.46 | 1235.97 | 1131.98 | 1157.37 | 1157.369995 | 76349800000 |

The data used in this paper will all be adjusted closing price because it eliminates the need to adjust stock pricing for splits and dividends, thus making the analysis of the data accurate.

A split is when a company decides to split each share into multiple shares, each at a lower value, thus making the stock more marketable. For example, if one share of stock Z is worth $300. The company decides to split the stock one hundred-for-one. Now the company has 100 times more shares available to trade, each at a price of $3. Nevertheless, if an investor owned 1 share before the split, he would own 100 shares after the split and therefore maintains the value of his investment.

A dividend is when a company distributes part of its profits back to its investors, which in turn lowers their stock price by the same amount. For example, stock B declares a $5 cash dividend and is trading at $105 dollars per share before the dividend date. On the dividend date, the stock price is reduced by $5, and the adjusted closing price becomes $100.

When testing investing models and strategies, the graph will show two lines. One is defined as continuous investing which represents buying a stock at the beginning of the testing period and holding it, making no changes or changes in position. The other is indicative investing which represents investing using the indicators given by the model or strategy being utilized.

### 4.2. Volatility

Volatility is defined as the uncertainty or risk related to the size of changes in a security's value or the statistical measure of the dispersion of returns for a given. Standard deviation is used as a statistical measurement in finance that sheds light on the volatility of an investment. The greater the standard deviation, the greater the variance between each dates price and the mean and a larger price range, therefore a riskier stock.

The standard deviation formula is based on the average deviation, the average of the distance from each data value to the mean.

$$average\ deviation = \frac{|x_1 - \bar{x}| + |x_2 - \bar{x}| + \cdots + |x_n - \bar{x}|}{n}$$

In order to eliminate absolute values which, become difficult to work with larger numbers and datasets, let's square the numerator. This provides us with the variance formula which measures the squared deviation. To make this deviation just square root it.

$$variance = \frac{(x_1 - \bar{x})^2 + (x_1 - \bar{x})^2 + \cdots + (x_1 - \bar{x})^2}{n}$$

$$= \sqrt{\frac{(x_1 - \bar{x})^2 + (x_1 - \bar{x})^2 + \cdots + (x_1 - \bar{x})^2}{n}}$$

This formula can be rewritten as a summation function instead of the numerator.

$$= \sqrt{\frac{\Sigma(x - \bar{x})^2}{n}}$$

Now if you have ever seen the standard deviation formula you may ask why the denominator above is n and not n-1. This is what is called the degrees of freedom which is meant to show that if you do not have a final value in the dataset it is possible to find it. If we know the mean of the dataset beforehand you only need n-1 data points and the final value can be figured out.

$$\sigma = \sqrt{\frac{\Sigma(x - \bar{x})^2}{n - 1}}$$

Let's look at some examples of how the volatility can be described using the standard deviation.

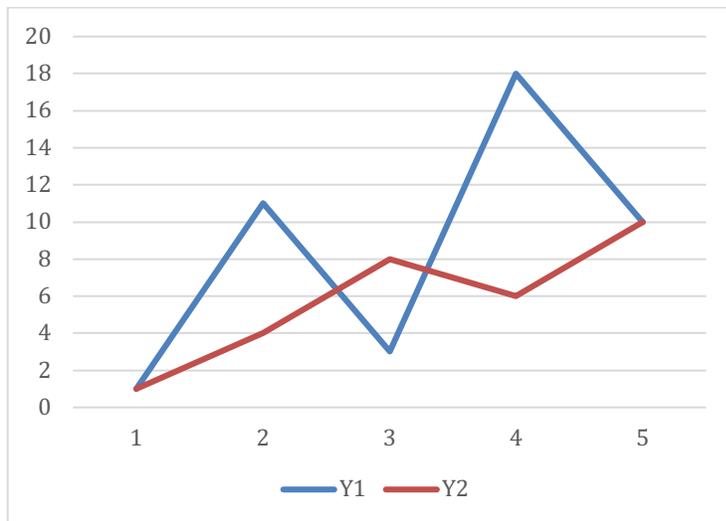

| X | $Y_1$ | $Y_2$ |
|---|---|---|
| 1 | 1 | 1 |
| 2 | 11 | 3 |
| 3 | 3 | 7 |
| 4 | 20 | 6 |
| 5 | 10 | 10 |

Over the period of the 5 trading days, both stocks began at the same point and ended at the same point, 900% increase. The difference is visible however, the blue line ($Y_1$) was more volatile, experiencing larger and more frequent jumps and dives in price. This is reflected in the standard deviation in order to accurately reflect the volatility. The more volatile stock price will have a higher standard deviation.

$$\sigma(Y_1) = 6.72309$$
$$\sigma(Y_2) = 3.136877428$$

### 4.3. K-Nearest Neighbor

The KNN algorithm is one that uses and stores all available data and the classifies the new data or case based on a similarity measure or Euclidean distance on a graph. A classification phase is undergone in which k is defined by a user and each new data point is assigned a label which is most frequent amongst the k existing data points nearest to it.

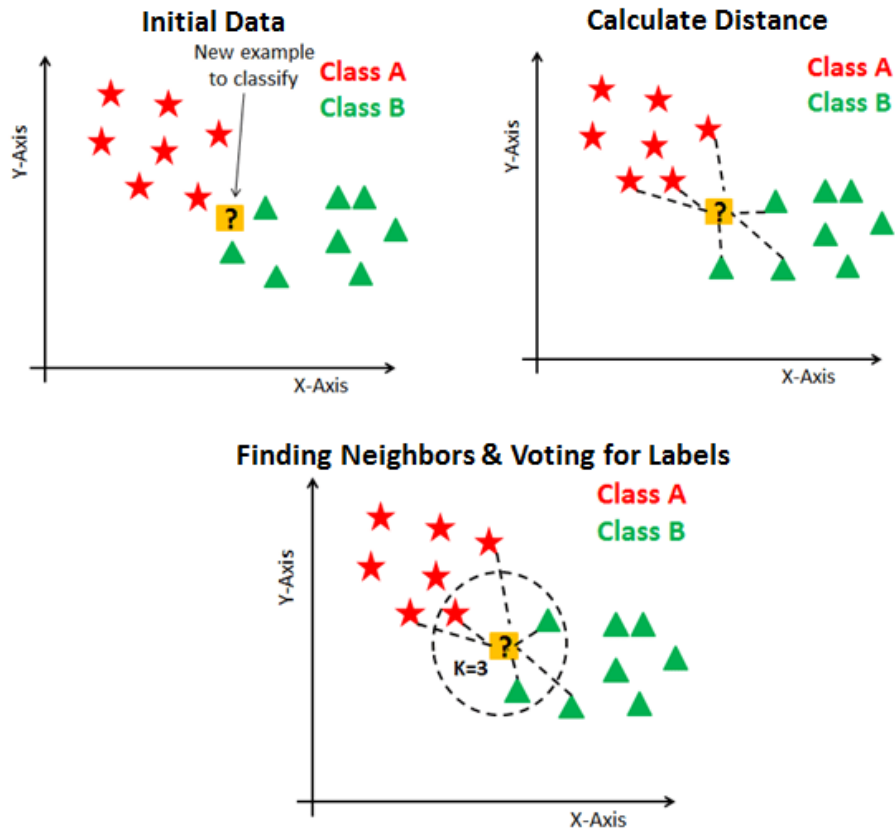

## 4.4. Logistic Regression

Regression takes a dataset and tries to find a mathematical relationship/formula relating the x variable to the y variable. Let's use an example, a baker sells donuts, x for a dozen, and makes y dollars in profit. He continues to increase the price but at a certain point, his profits go down because now he charges too much for a dozen donuts and people don't want to buy them from him anymore. How can he find the highest price he can charge without turning costumers away with a high price tag? Let's use the following data to try to predict that value.

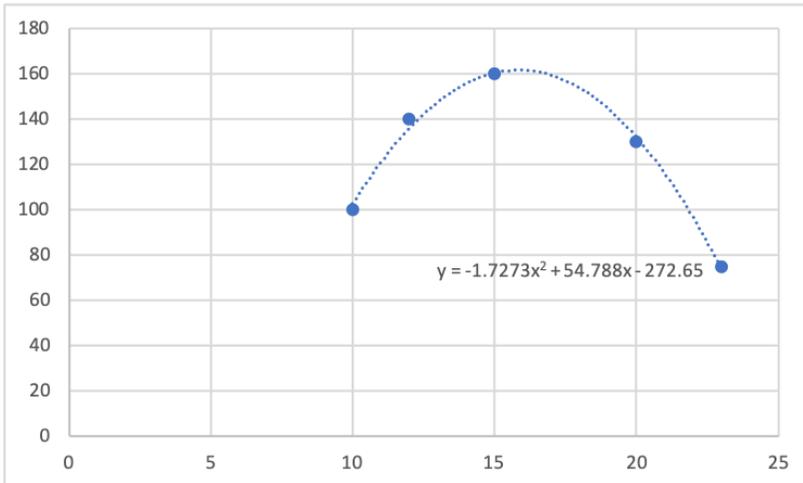

| Price per dozen donuts | Profit |
|---|---|
| 10 | 100 |
| 12 | 140 |
| 15 | 160 |
| 20 | 130 |
| 23 | 75 |

Plotting this on a graph we can use regression to try to estimate the highest point before which the profits begin to decrease. Since we know that it increases then decreases, we can use quadratic regression and try to apply as close a formula as possible to predict the peak of the parabola, which would theoretically maximize profits. The formula that best fits the graph is $y = -1.7273x^2 + 54.788x - 272.65$. Using this formula, we can find the x value that provides the largest y value (profit), which is around (15.86, 161.803), meaning if the baker charges 15.86 dollars per dozen he will maximize his profits at $161.803.

### 4.5. Moving Average Crossovers

Moving averages are a widely used indicator for the analysis of stocks that functions to filter out noise and show general trends for chosen periods of time. Moving averages are considered lagging indicators because it is based on past prices and, by itself, only reflects what has happened in the past.

There are three main types of moving averages, the simple moving average, the weighted moving average, and the exponential moving average. We will be using the simple moving average (SMA) due to its easy calculation and versatile application. The SMA is calculated by taking the average of a stock over a defined time period.

$$Simple\ Moving\ Average (SMA) = \frac{A_1 + A_2 + A_3 + \cdots + A_n}{length\ of\ period(n)}$$

Let's suppose this is a data table showing the movement of stock A.

| Date | Adj. Closing of Stock A | 5-day SMA |
| --- | --- | --- |
| Jan 1 | 20 | N/A |
| Jan 2 | 22 | N/A |
| Jan 3 | 24 | N/A |
| Jan 4 | 25 | N/A |
| Jan 5 | 23 | $\frac{20 + 22 + 24 + 25 + 23}{5} = 22.8$ |
| Jan 6 | 22 | $\frac{22 + 24 + 25 + 23 + 22}{5} = 23.2$ |
| Jan 7 | 27 | $\frac{24 + 25 + 23 + 22 + 27}{5} = 24.2$ |

When two moving averages, one long and one short, are graphed, trends can be predicted using the crossover of the moving averages. If the short SMA crosses the long SMA moving upwards this indicated a good entry point and when it crosses back below the long SMA it predicts a strong downward trend, therefore, indicating an exit.

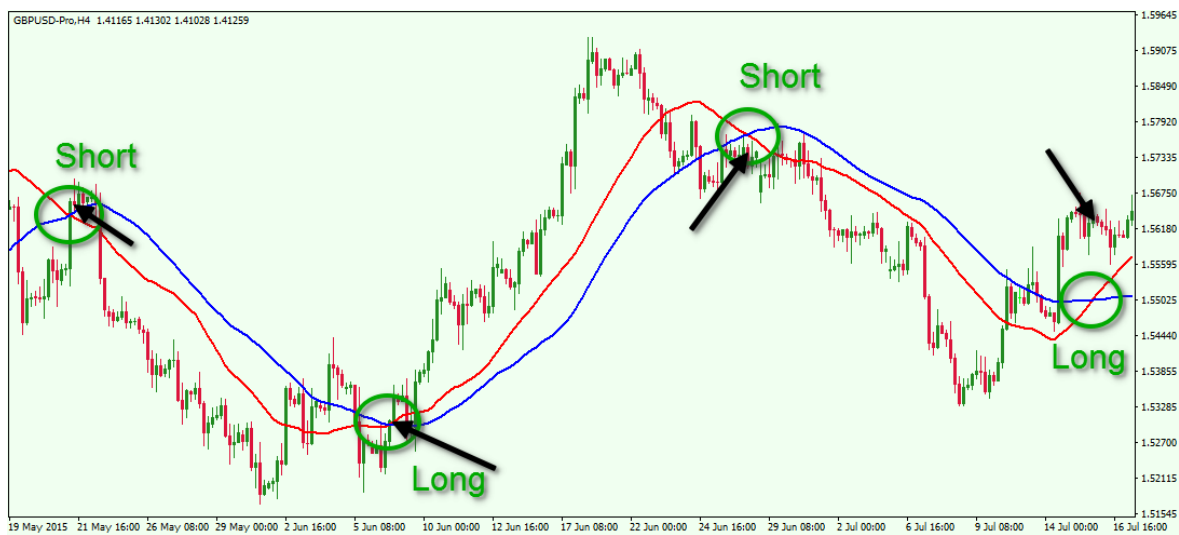

## 5. Investigation

### 5.1. Regression

#### 5.1.1. Program Development

To create a linear regression model, we will have to do a couple of things first. We will need to gather user input for a selected stock, a beginning date and end date for training data, and then an end date for testing data. Then we must split up the data into two different data frames, one for creating and testing the model, and one for using the model to invest.

```
14.  ticker_input = input("Please enter the stock tickers you would like to invest in.\n")
15.
16.  start_date = input("Here enter the date from which data will begin to be taken from (Please put it
     into the format of YYYY-MM-DD)\n")
17.
18.  test_date = input("Here enter the date from which training and testing will end (Please put it int
     o the format of YYYY-MM-DD)\n")
19.
20.  end_date = input("Here enter the date at which we will stop taking data from (Please put it into t
     he format of YYYY-MM-DD)\n")
21.
22.  data = pdr.DataReader(ticker_input, test_date, end_date)#['Adj Close']
23.  indices = data.index.tolist()
24.  test_date = indices[0]
25.  test_data = pdr.DataReader(ticker_input, start_date, test_date)
```

After downloading data, it must be split up at a ratio of 80/20 for training data and testing data for the X and Y separately. This data will then be used to fit the model and to test its confidence.

```python
47.  #This Forecast Out will predict 1 day into the future
48.  f_out = int(1)
49.  test_data['Predicting'] = test_data['Adj Close'].shift(-f_out)
50.
51.  X = np.array(test_data.drop(['Predicting'],1))
52.  X = pp.scale(X)
53.  X_prediction = X[-f_out:]
54.  X = X[:-f_out]
55.
56.  Y = np.array(test_data['Predicting'])
57.  Y = Y[:-f_out]
58.
59.  X_train, X_test, Y_train, Y_test = tts(X, Y, test_size = 0.2)
60.
61.  model = lr()
62.
63.  model.fit(X_train,Y_train)
64.
65.  #Prints confidence of initial model
66.  conf = model.score(X_test,Y_test)
67.  print("Model Confidence: ", conf)
```

Now to use this model when investing, we must use the predicted price and see if it predicts the stock will increase or decrease in value, and the program will invest accordingly. In addition, to ensure the validity of the test no data is used from past the predicted date's price until after it has been predicted at which point the actual movement of the stock is added into the training set and the model is refitted.

```python
94.       prediction = model.predict(X_prediction)
95.       indices = data.index.tolist()
96.       test_date = indices[r]
97.       return(prediction - test_data.iloc[len(test_data)-1]['Adj Close'])
105. for r in range (1,len(data)):
106.      x = investing_model(r)[0]
107.      if x > 0:
108.          last_true = last_true+(data.iloc[r]['Adj Close']-data.iloc[r-1]['Adj Close'])
109.      indicative_investing.append(last_true)
```

Finally, results are compared using the standard deviation to reflect volatility and the outperformance percentage of investing using the model.

```python
27.  def outperformance_percentage():
28.      outperformance_count = 0
```

```
29.        total_count = 0
30.        global comparison
31.
32.        for r in range (0, len(comparison.index)):
33.            if comparison[ticker_input + ' Indicative Investing'].iloc[r] > comparison[ticker_input +
    ' Continuous Investing'].iloc[r]:
34.                outperformance_count += 1
35.            total_count += 1
36.        outperformance_percentage = (outperformance_count / total_count) * 100
37.
38.        return outperformance_percentage
39.
40. def standard_deviation(data_list):
41.     new_list = data_list
42.     standard_deviation_sum = 0
43.     for i in range(0,len(new_list)):
44.         standard_deviation_sum += (new_list[i] - statistics.mean(new_list)) ** 2
45.     return(((standard_deviation_sum / (len(new_list) - 1)) ** (0.5)))
```

### 5.1.2. Testing

To test the model, we will invest into three of the most highly traded stocks (Advanced Micro Devices [AMD], General Electric [GE] and Bank of America [BAC]) as well as two major indexes (S&P 500 [SPY] and the Dow Jones Index [^DJI]) from Jan 1. 2017 to Jan 1. 2019. For training data, the model will use 2 years of historical data, Jan 1. 2015 to Jan 1. 2017.

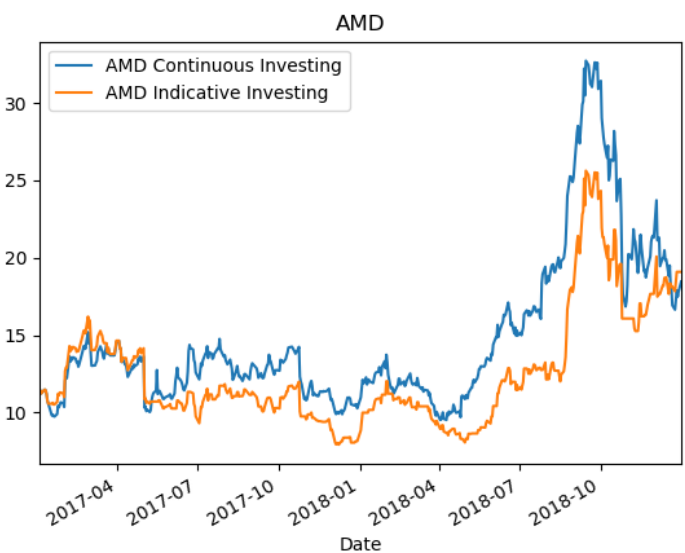

**AMD**

Continuous Investing Final Price: 18.46

Indicative Investing Final Price: 19.09

AVG Model Confidence: 0.994

Outperformance Percentage: 17.33%

Volatility Ratio: 0.7355

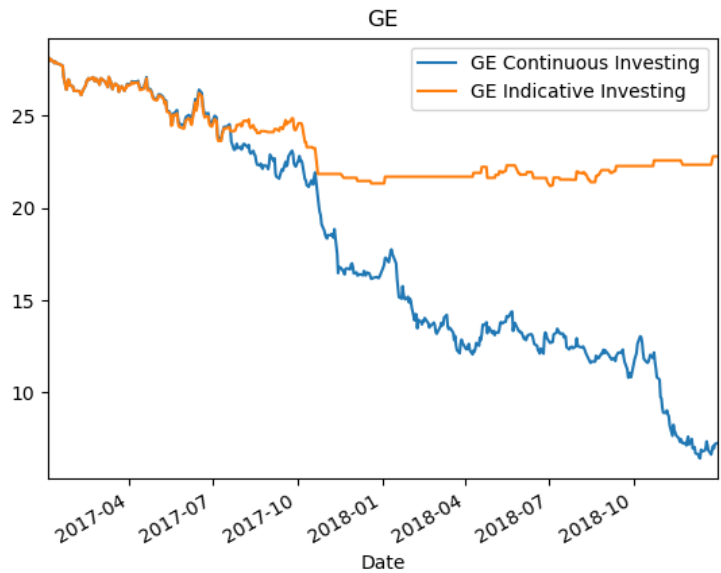

**GE**

Continuous Investing Final Price: 7.27

Indicative Investing Final Price: 22.79

AVG Model Confidence: 0.986

Outperformance Percentage: 72.51%

Volatility Ratio: 0.303

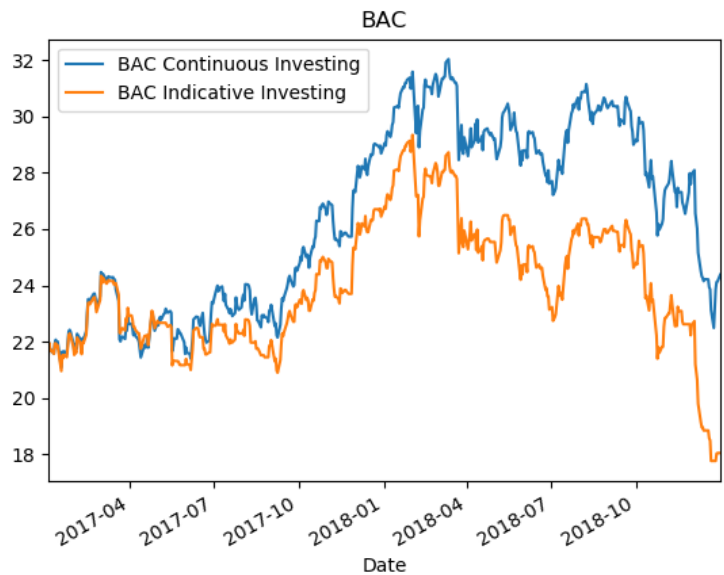

**BAC**

Continuous Investing Final Price: 24.38

Indicative Investing Final Price: 18.05

AVG Model Confidence: 0.979

Outperformance Percentage: 16.37%

Volatility Ratio: 0.696

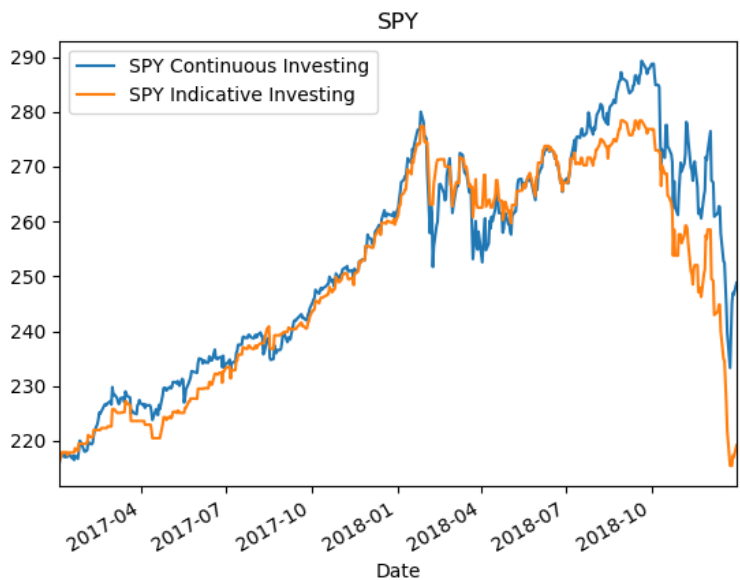

**SPY**

Continuous Investing Final Price: 248.82

Indicative Investing Final Price: 219.19

AVG Model Confidence: 0.974

Outperformance Percentage: 27.49%

Volatility Ratio: 0.97

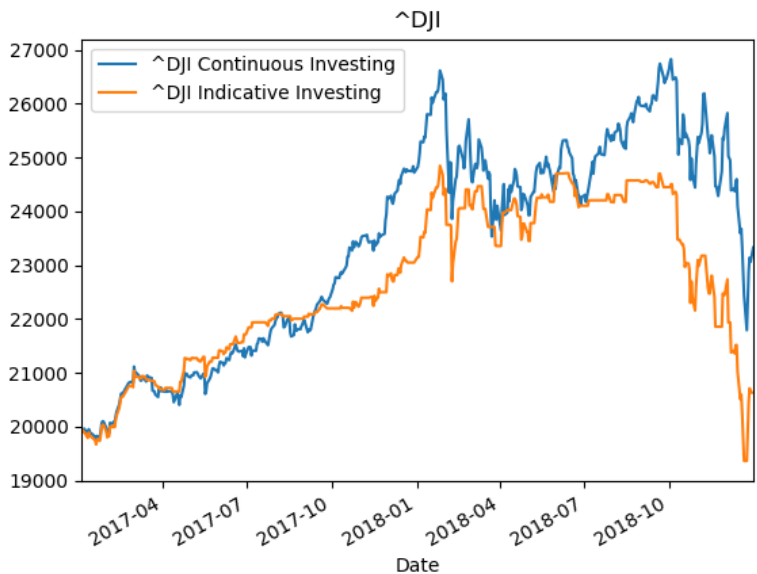

**^DJI**

Continuous Investing Final Price: 23327

Indicative Investing Final Price: 20633

AVG Model Confidence: 0.962

Outperformance Percentage: 25.49%

Volatility Ratio: 0.7335

### 5.2. K-Nearest Neighbor

#### 5.2.1. Program Development

Development for the KNN model program is very similar to the regression model program due to the convenience and versatility of the sklearn library. Our first step remains to find the ticker and dates that the user inputs and use that to download data.

The data is then split with an 80/20 split for training and testing data, after which a value is predicted, the program invests accordingly to anticipated returns being either positive or negative, adds the actual stock price to training data, and refits the model for the next iteration.

The only difference in programming is the change in model from LinearRegression() to KneighborsRegressor(). In addition, the data must be scaled accordingly, and the k must be input, although GridSearchCV optimizes this choice within a certain set of set parameters.

```
49.  #This Forecast Out will predict 1 day into the future
50.  f_out = int(1)
51.  test_data['Predicting'] = test_data['Adj Close'].shift(-f_out)
52.
53.  X = np.array(test_data.drop(['Predicting'],1))
54.  X = pp.scale(X)
55.  X_prediction = X[-f_out:]
56.  X = X[:-f_out]
57.
58.  Y = np.array(test_data['Predicting'])
59.  Y = Y[:-f_out]
60.
61.  X_train, X_test, Y_train, Y_test = tts(X, Y, test_size = 0.2)
62.
63.  X_train_scaled = scaler.fit_transform(X_train)
64.  X_test_scaled = scaler.fit_transform(X_test)
65.
66.  parameters = {'n_neighbors':[2,3,4,5,6,7,8,9,10,11,12,13,14,15]}
67.
68.  #define KNN algorithm and k value (k=5)
69.  knn= KNeighborsRegressor()
70.  model = GridSearchCV(knn, parameters, cv = 5)
71.  model.fit(X_train,Y_train)
```

### 5.2.2. Testing

To test the model, we will invest into three of the most highly traded stocks (Advanced Micro Devices [AMD], General Electric [GE] and Bank of America [BAC]) as well as two major indexes (S&P 500 [SPY] and the Dow Jones Index [^DJI]) from Jan 1. 2017 to Jan 1. 2019. For training data, the model will use 2 years of historical data, Jan 1. 2015 to Jan 1. 2017.

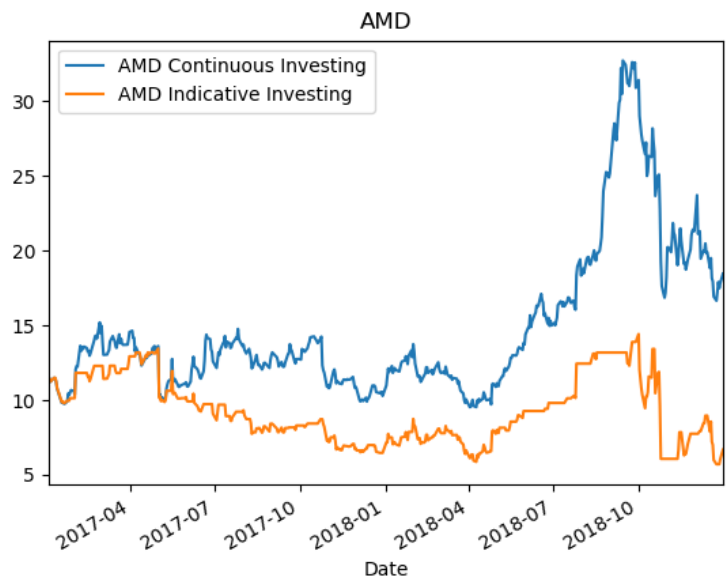

**AMD**

Continuous Investing Final Price: 18.46

Indicative Investing Final Price: 6.67

AVG Model Confidence: 0.987

Outperformance Percentage 3.78%

Volatility Ratio: 0.43

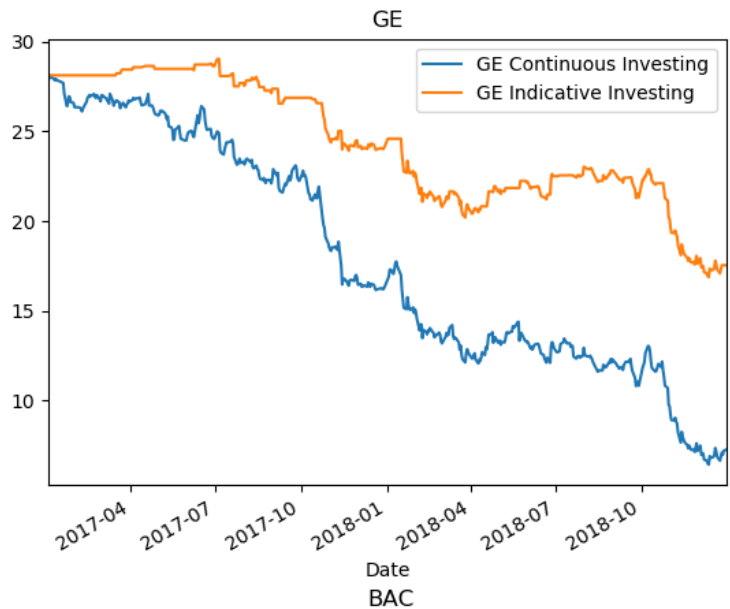

**GE**

Continuous Investing Final Price: 7.27

Indicative Investing Final Price: 17.54

AVG Model Confidence: 0.9817

Outperformance Percentage: 99.6%

Volatility Ratio: 0.528

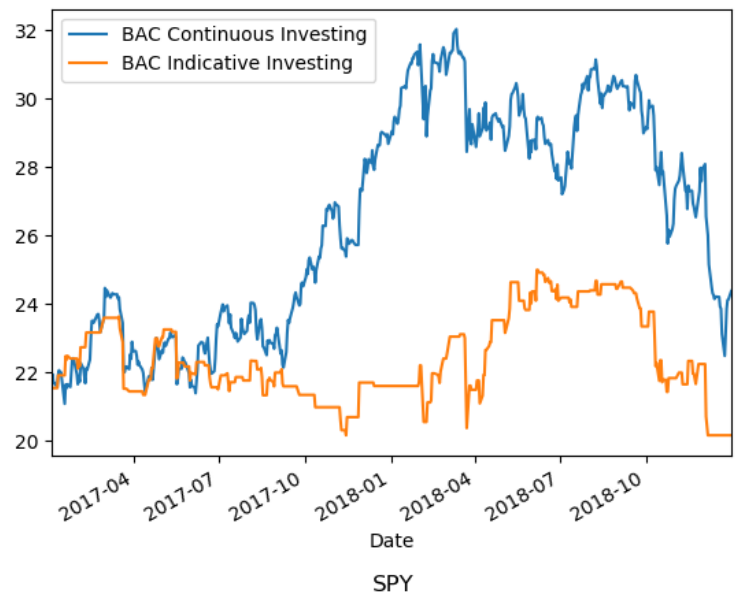

**BAC**

Continuous Investing Final Price: 24.38

Indicative Investing Final Price: 20.17

AVG Model Confidence: 0.981

Outperformance Percentage: 9.96%

Volatility Ratio: 0.383

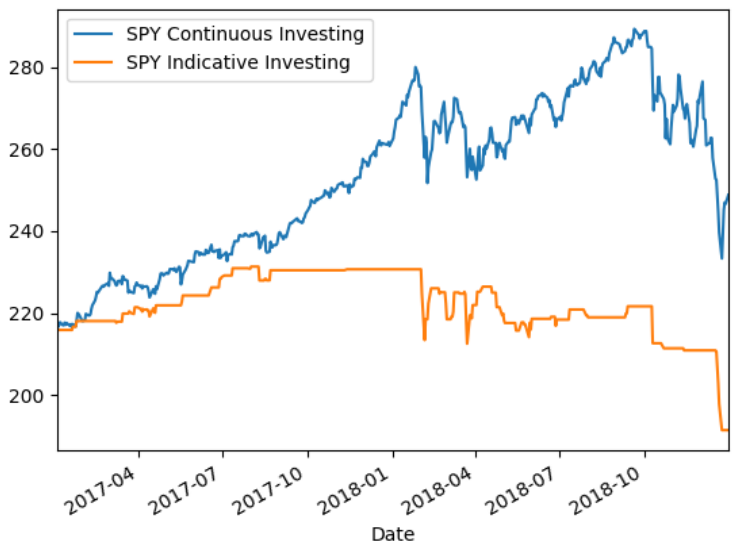

**SPY**

Continuous Investing Final Price: 248.82

Indicative Investing Final Price: 191.45

AVG Model Confidence: 0.948

Outperformance Percentage: 0%

Volatility Ratio: 0.355

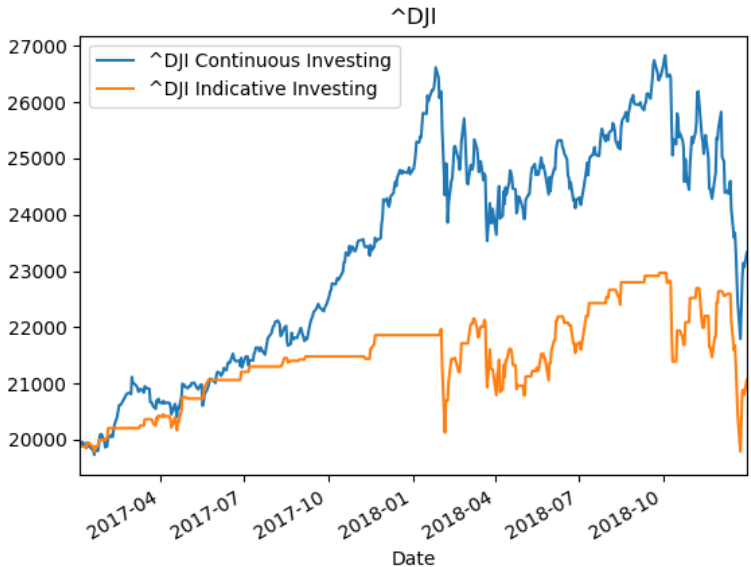

**^DJI**

Continuous Investing Final Price: 23327

Indicative Investing Final Price: 21064

AVG Model Confidence: 0.954

Outperformance Percentage: 4.98%

Volatility Ratio: 0.397

### 5.3. Moving Average Crossover

#### 5.3.1. Program Development

To use moving averages to indicate times to invest, we will have to write a python program that will calculate the moving average with a user input of a stock, dates, and the moving average periods. We have to set up an input for the user that includes ticker symbol for the stock we will be investing in, dates for the data, and two numerical inputs for the periods, one of which would be longer than the other.

```
110. ticker_input = input("Please enter the stock ticker you have chosen to invest in. \n")
111.
112. start_date= input("Here enter the date from which data will begin to be taken from (Please put it
     into format of YYYY-MM-DD)\n")
113.
114. end_date= input("Here enter the date at which we will stop taking data from (Please put it into fo
     rmat of YYYY-MM-DD)\n")
115.
116. ma_list = []
117.
118. ma_value = int(input("What would you like the 1st period to be for your moving average?\n"))
119. ma_list.append(ma_value)
120. ma_value = int(input("What would you like the 2nd period to be for your moving average?\n"))
121. ma_list.append(ma_value)
```

Then using the formula for SMA, we can define a function that will calculate it given a starting day that is the last day in the period(start), and the length of the period itself.

$$Simple\ Moving\ Average (SMA) = \frac{A_1 + A_2 + A_3 + \cdots + A_n}{length\ of\ period(n)}$$

```
31. def simple_moving_average(length,start):
32.     s= 0
33.     for i in range (0,int(length)):
34.         #print (ticker_data[start-i])
35.         s += ticker_data[start-i]
36.     return (s/length)
```

We then create a new data frame to hold the moving averages separate from the stock data itself. For each of the input moving average period values, we will need to calculate every single moving average in between the data input for the dates. Then upon a deeper inspection of the data we see that with a period of n, we will have no values for the first n-1 data points so we must crop all data frames so that we start with viable data and not blank points.

```
59. moving_averages = pd.DataFrame(index = ticker_data.index)
60. for i in ma_list:
61.     list2 = []
62.     for v in range(0,len(ticker_data.index)):
63.         if v < i-1:
64.             list2.append(0)
65.         if v > i-1 or v == i-1:
66.             list2.append (simple_moving_average(i,v))
67.     moving_averages[ticker_input + " "+str(i) +" Day Moving Average"] = list2
68.
69. #Remove all values that are NA and makes sure that all lists start from the same date
70. ticker_data = ticker_data.iloc[max(ma_list)-1:]
71. moving_averages = moving_averages.iloc[max(ma_list)-1:]
```

When the calculation and graphing of the moving averages was complete, there was an issue of how to tell when the moving averages cross. It's unlikely that they would meet at a single point but rather the lines intersect between points, thus making it impossible to set them equal to each other and solving. To do this we found the differences between the short-term trend and the long-term trend, and whenever in between a data point and the one that preceded it had different signs we know that the lines crossed over. This is now used to indicate times to buy, sell, or continue with whatever is happening. When the short term, crosses over the long term and becomes higher than it, it is time to buy, and we will continue holding the position until the short-term average falls below the long-term average, and vice-a-versa.

```python
74. Calculates the difference in value between moving averages for every data-point
75. def difference_calc(x):
76.     return x[1]- x[0]
77.
78. differences=pd.DataFrame(index = ticker_data.index)
79.
80. differences = moving_averages.apply(difference_calc,axis=1)
81.
82. #Tests for sign differences between moving averages to sense crossovers
83. def test(x):
84.     if differences[x] < 0 and differences[x-1] > 0:
85.         return "sell"
86.     if differences [x] > 0 and differences[x-1] < 0:
87.         return "buy"
88.     if differences [x] > 0 and differences[x-1] > 0:
89.         return "continue_s"
90.     if differences [x] < 0 and differences[x-1] < 0:
91.         return "continue_b"
```

We develop an initial start position based on whether or not the short term SMA is above the long term SMA. Then for each data point we evaluate the existing position and the indication from sign of the differences to dictate our change from buy to sell or lack of change in position.

```python
98. if differences[0] > 0:
99.     position = "bought"
100.    if differences[0] < 0:
101.        position = "sold"
102.
103.    for r in range (0,len(ticker_data.index)):
104.        if position == "bought" and test(r) == "continue_b":
105.            last_true = last_true+(ticker_data[r]-ticker_data[r-1])
106.        if position == "bought" and test(r) == "sell":
107.            last_true = last_true+(ticker_data[r]-ticker_data[r-1])
108.            position == "sold"
109.        if position == "sold" and test(r) == "buy":
110.            position = "bought"
111.        indicative_investing.append(last_true)
```

Then we plot together the final price of our portfolio if we were to invest continuously together with the price of our portfolio if we had used the SMA crossover strategy to see the benefits of holding a stock for a long period of time versus strategically buying and selling to anticipate changes in trends.

```
113.            comparison = pd.DataFrame(index = ticker_data.index)
114.            comparison[ticker_input + ' Continuous Investing'] = continuous_investing
115.            comparison[ticker_input + ' Indicative Investing'] = indicative_investing
```

Finally, to be able to compare the investing strategy more accurately we can create functions to compare the volatility and the percentage of times that when the market closed, using the moving average crossover strategy outperformed a continuous hold of the stock.

```
38. def standard_deviation(data_list):
39.     new_list = data_list
40.     standard_deviation_sum = 0
41.     for i in range(0,len(new_list)):
42.         standard_deviation_sum += (new_list[i] - statistics.mean(new_list)) ** 2
43.     return(((standard_deviation_sum / (len(new_list) - 1)) ** (0.5)))
44.
45. def outperformance_percentage():
46.     outperformance_count = 0
47.     total_count = 0
48.     global comparison
49.
50.     for r in range (0, len(comparison.index)):
51.         if comparison[ticker_input + ' Indicative Investing'].iloc[r] > comparison[ticker_input + ' Continuous Investing'].iloc[r]:
52.             outperformance_count += 1
53.         total_count += 1
54.     outperformance_percentage = (outperformance_count / total_count) * 100
55.
56.     return outperformance_percentage
```

### 5.3.2. Determining Optimal Moving Average Periods

When beginning to do investing simulations to test the moving average program, I began to play around with the moving average periods. With so many choices and each combination providing different results, I naturally wondered which combination generally provides the best results. I created a program which would test the results for using the crossover strategy with an input stock for every combination of Short SMA between 5 and 49 and Long SMA between 10 and 149 (Long SMA must be larger than Short SMA). This test would record the percent of market close in which using the strategy would provide more profit than continuous investing, and the ratio of volatility of the crossover strategy investing and continuous investing. The following 30 stocks were tested for the last 5 years (from Jan 1, 2014 to Jan 1, 2019):

| AAPL | AMD | BABA | BBY | DAL | DIS |
| --- | --- | --- | --- | --- | --- |
| DWDP | EBAY | FB | FL | GE | GIS |
| GOOGL | HPQ | INTC | JNJ | KSS | M |
| MET | NFLX | NOK | NVDA | NXPI | QCOM |
| S | SBUX | SIRI | T | TEVA | TSLA |

This test consisted of 169,880 trials and 339,760 data points. There are some visible trends in the data and periods that on average provide both the greatest profit and/or least volatility.

The first visible trends were seen after an analysis of average outperformance percentages and volatilities of the short SMA's alone. The smaller the shorter SMA is the more volatile the investment is, but the better it performs as well. Almost the exact same trend can be seen with the long SMA.

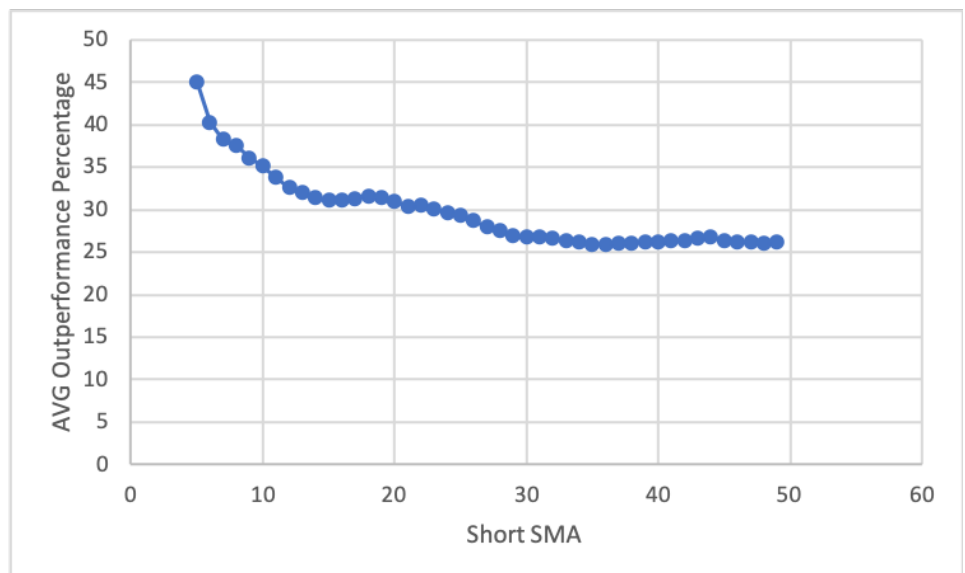

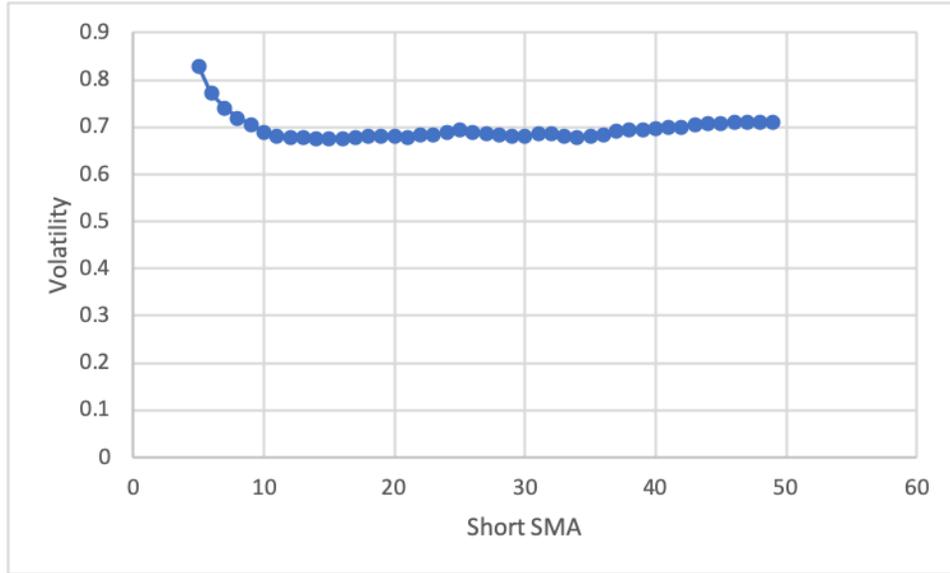

There were no visible trends when analyzing the ratio of short SMA to long SMA when comparing them to volatility or performance, which shows that optimal periods for the strategy have specific values rather than a ratio in which the long SMA period must be double the short SMA period.

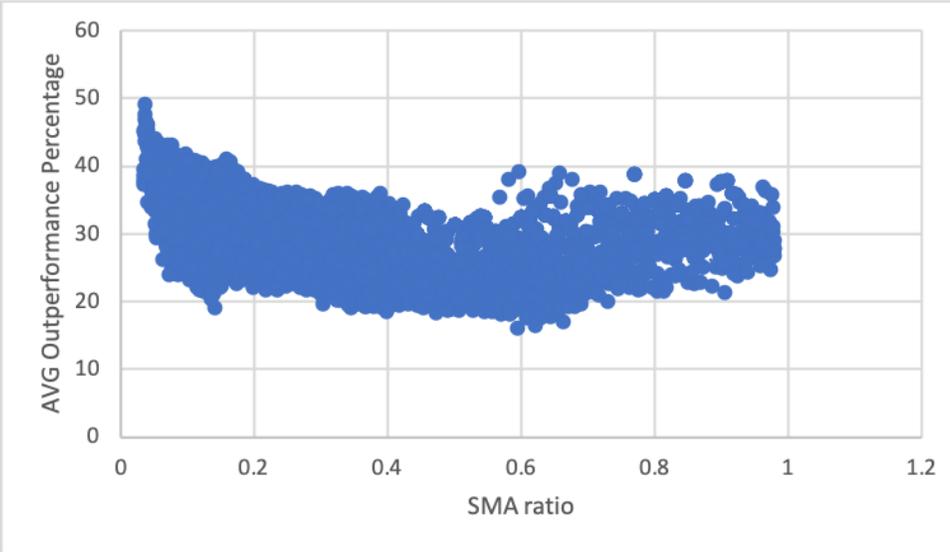

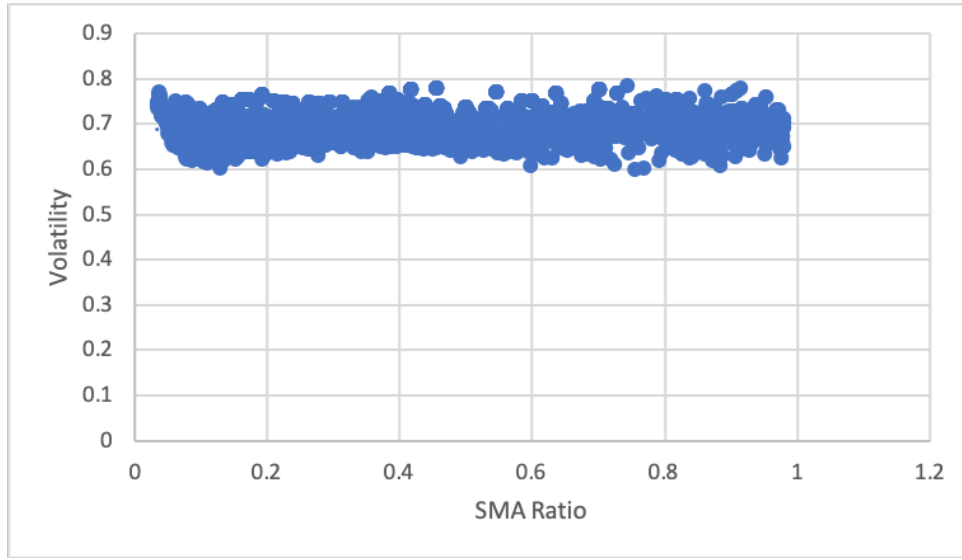

When taking each average volatility and average outperformance percentage for each pair and plotting them against each other there is a clear trend in the data. It's visible that better performance comes from investments that are more volatile, further confirming the notion of high risk, high reward.

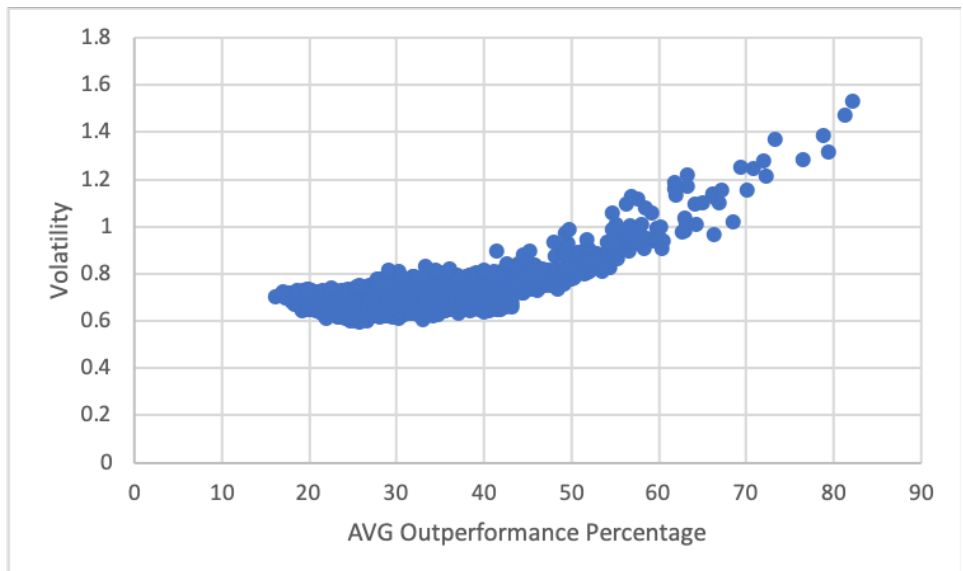

Finally, below is the list of the top 5 and bottom 5 pairs of moving average periods to use with moving average crossovers with regards to percent outperformance of continuous investing and volatility.

Average Outperformance Percentage

| Rank | Short | Long | Outperformance % |
|---|---|---|---|
| 1 | 5 | 10 | 82.18101802 |
| 2 | 5 | 11 | 81.3443281 |
| 3 | 5 | 13 | 79.32189281 |
| 4 | 5 | 12 | 78.84323528 |
| 5 | 5 | 14 | 76.52557515 |
| 5476 | 48 | 78 | 17.3441826 |
| 5477 | 49 | 74 | 17.07247726 |
| 5478 | 49 | 77 | 16.97714385 |
| 5479 | 49 | 79 | 16.49604221 |
| 5480 | 47 | 79 | 16.16093988 |

Average Volatility

| Rank | Short | Long | Volatility |
|---|---|---|---|
| 1 | 33 | 44 | 0.59252 |
| 2 | 34 | 45 | 0.600884 |
| 3 | 33 | 43 | 0.60243 |
| 4 | 34 | 44 | 0.602612 |
| 5 | 11 | 86 | 0.603649 |
| 5476 | 5 | 13 | 1.317485 |
| 5477 | 6 | 10 | 1.371095 |
| 5478 | 5 | 12 | 1.383828 |
| 5479 | 5 | 11 | 1.469967 |
| 5480 | 5 | 10 | 1.533497 |

### 5.3.3. Testing

To test the model, we will invest into three of the most highly traded stocks (Advanced Micro Devices [AMD], General Electric [GE] and Bank of America [BAC]) as well as two major indexes (S&P 500 [SPY] and the Dow Jones Index [^DJI]) from Jan 1. 2017 to Jan 1. 2019. For training data, the model will use 2 years of historical data, Jan 1. 2015 to Jan 1. 2017.

# AMD

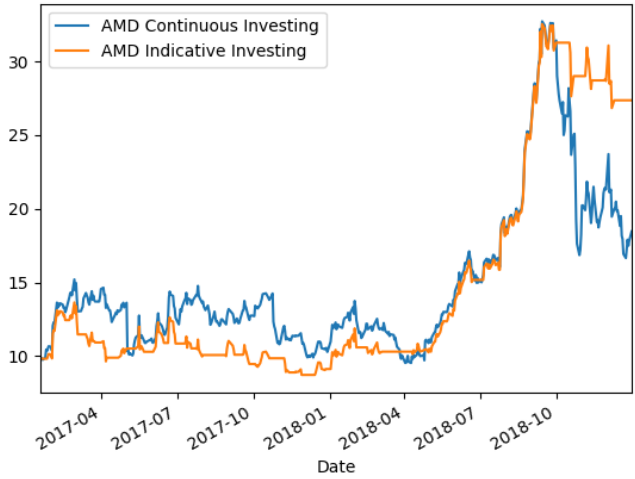 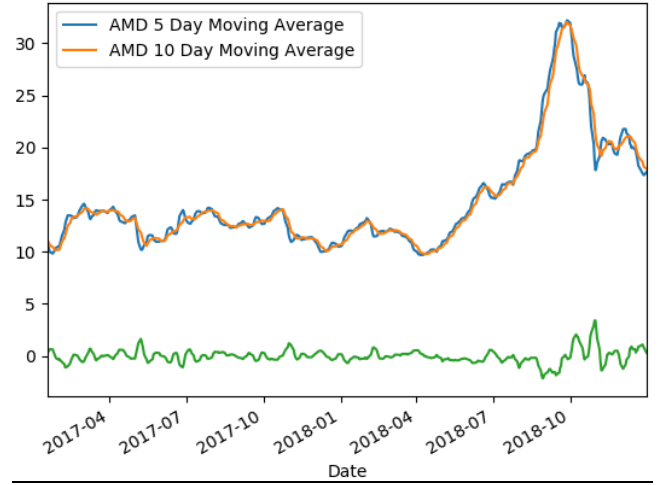

Continuous Investing Final Price: 18.46
Indicative Investing Final Price: 27.36
Outperformance Percentage: 19.47%
Volatility Ratio: 1.413

# GE

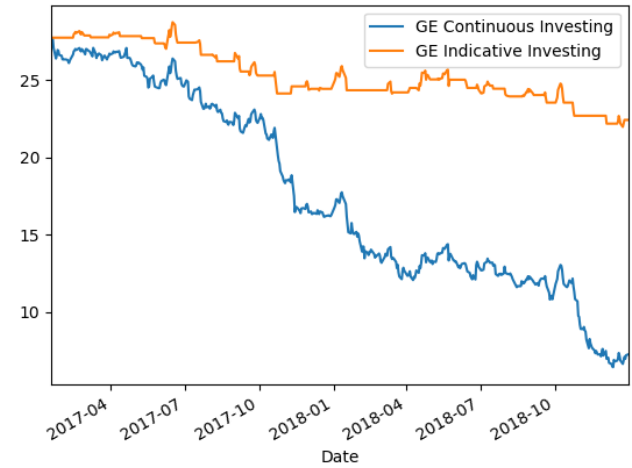 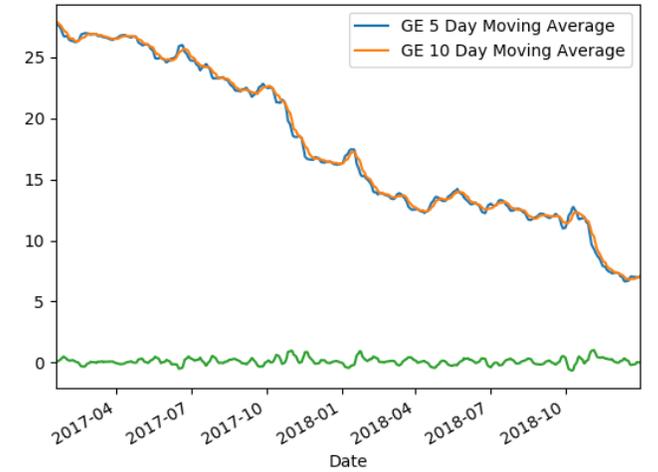

Continuous Investing Final Price: 7.27
Indicative Investing Final Price: 22.43
Outperformance Percentage: 99.797%
Volatility Ratio: 0.269

## **BAC**

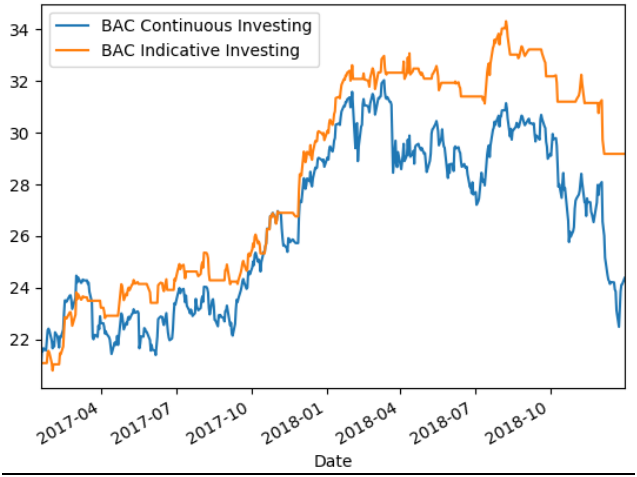 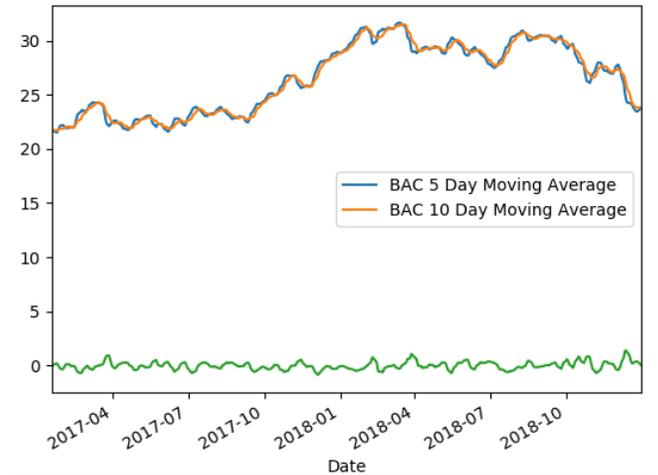

Continuous Investing Final Price: 24.38
Indicative Investing Final Price: 29.18
Outperformance Percentage: 88.23%
Volatility Ratio: 1.238

## **SPY**

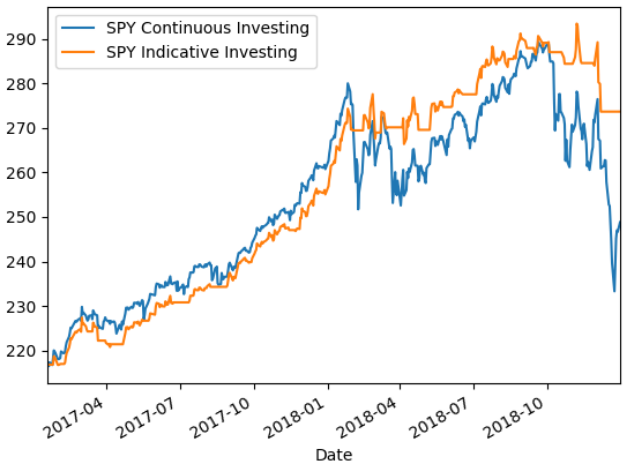 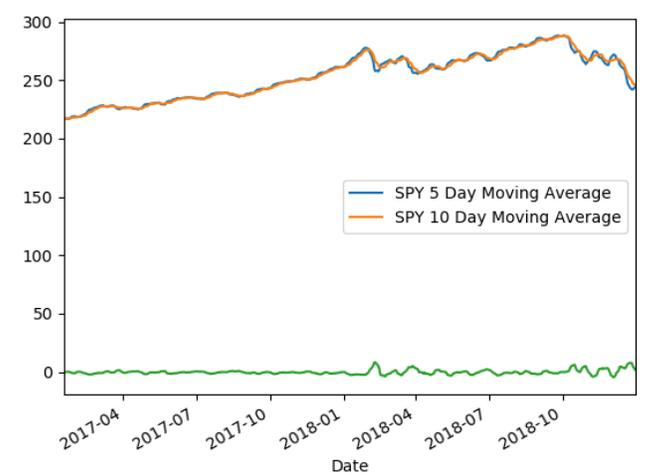

Continuous Investing Final Price: 248.82
Indicative Investing Final Price: 273.65
Outperformance Percentage: 46.85%
Volatility Ratio: 1.235

**^DJI**

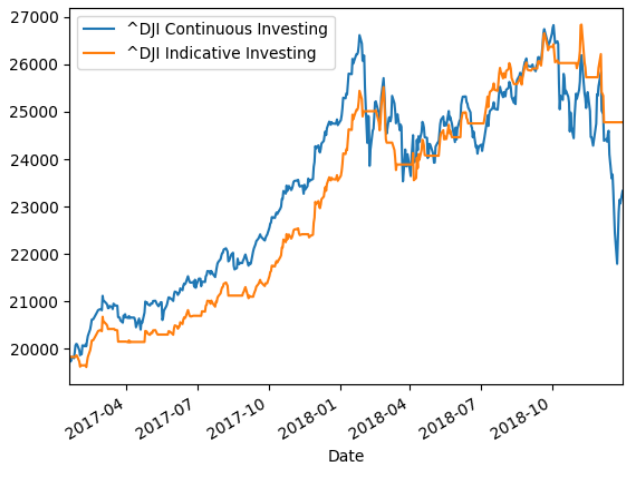 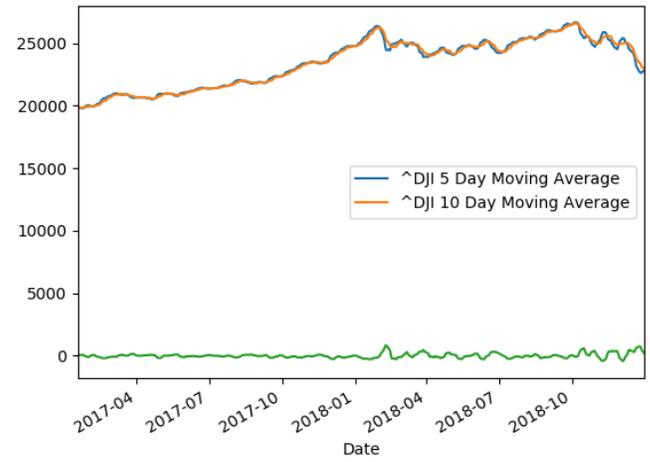

Continuous Investing Final Price: 23327.46
Indicative Investing Final Price: 24778.38
Outperformance Percentage: 24.138
Volatility Ratio: 1.132

## 6. Conclusions

Initially, we explored three models for investing. The linear regression and KNN model did not improve investment outcomes on the indexes and had varied results for the stocks. GE's drop was avoided with both machine learning models although implementing those models with BAC and AMD did not improve performance but did result in a less volatile investment.

The moving average crossover strategy performed much better. It outperformed continuous investing every time however for most trials it was more volatile. The results presented in **5.3.3.** were achieved by using a 5-day SMA and 10-day SMA as the moving averages. This was proved to be the most effective combination of periods to use in order to optimize performance in **5.3.2.**

**5.3.2.** consists of 169,880 trials which when analyzed show that the following combinations of periods are the best and worst to use for increasing performance and decreasing volatility.

Average Outperformance Percentage

| Rank | Short | Long | Outperformance % |
|---|---|---|---|
| 1 | 5 | 10 | 82.18101802 |
| 2 | 5 | 11 | 81.3443281 |
| 3 | 5 | 13 | 79.32189281 |
| 4 | 5 | 12 | 78.84323528 |
| 5 | 5 | 14 | 76.52557515 |
| 5476 | 48 | 78 | 17.3441826 |
| 5477 | 49 | 74 | 17.07247726 |
| 5478 | 49 | 77 | 16.97714385 |
| 5479 | 49 | 79 | 16.49604221 |
| 5480 | 47 | 79 | 16.16093988 |

Average Volatility

| Rank | Short | Long | Volatility |
|---|---|---|---|
| 1 | 33 | 44 | 0.59252 |
| 2 | 34 | 45 | 0.600884 |
| 3 | 33 | 43 | 0.60243 |
| 4 | 34 | 44 | 0.602612 |
| 5 | 11 | 86 | 0.603649 |
| 5476 | 5 | 13 | 1.317485 |
| 5477 | 6 | 10 | 1.371095 |
| 5478 | 5 | 12 | 1.383828 |
| 5479 | 5 | 11 | 1.469967 |
| 5480 | 5 | 10 | 1.533497 |

On top of determining the optimal combination of SMA periods, several trends are evident as well. The smaller the short SMA period, the better it performs, and the more volatile it is in relation to continuous investing.

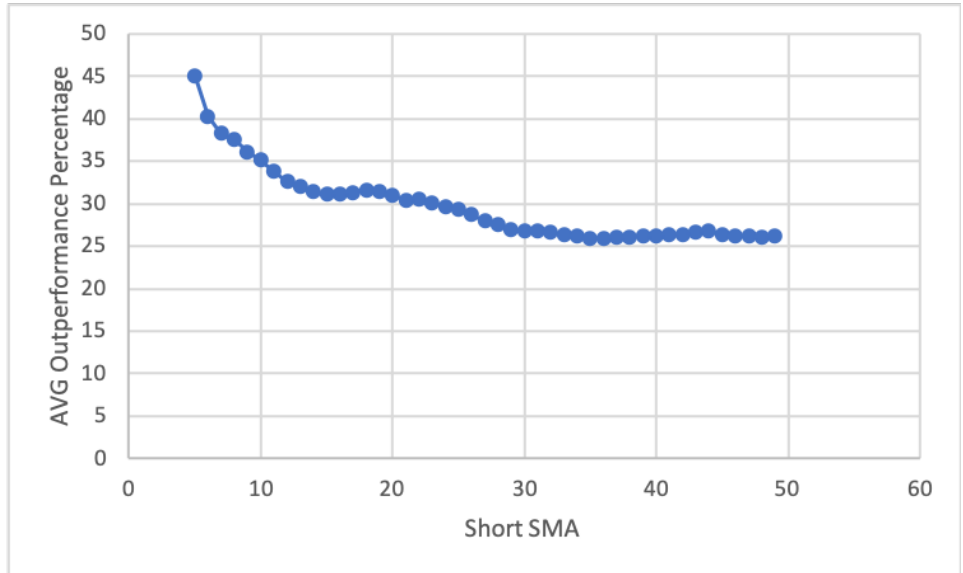

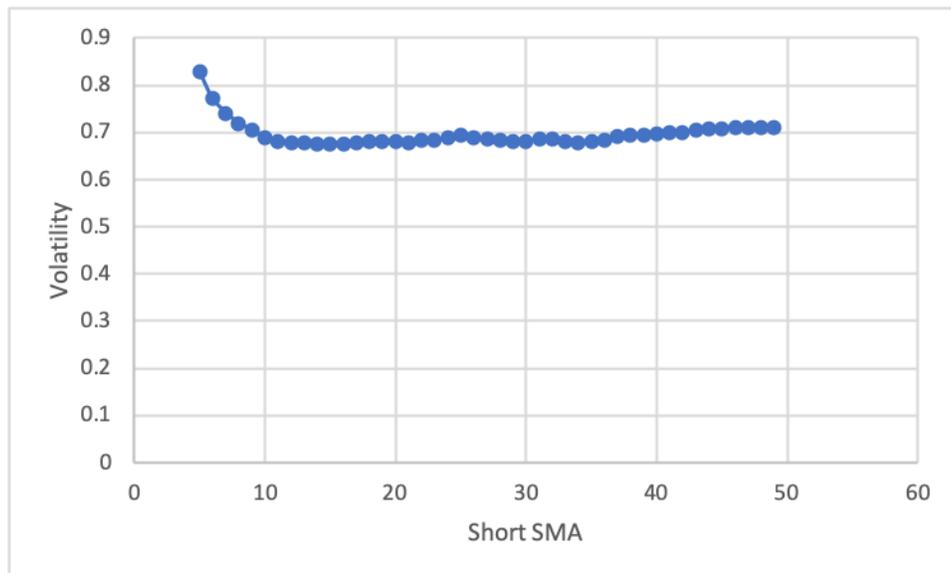

When comparing outperformance percentage and volatility, it is visible that they have a direct relationship. On average, trials where the investment becomes more volatile, it also performs better.

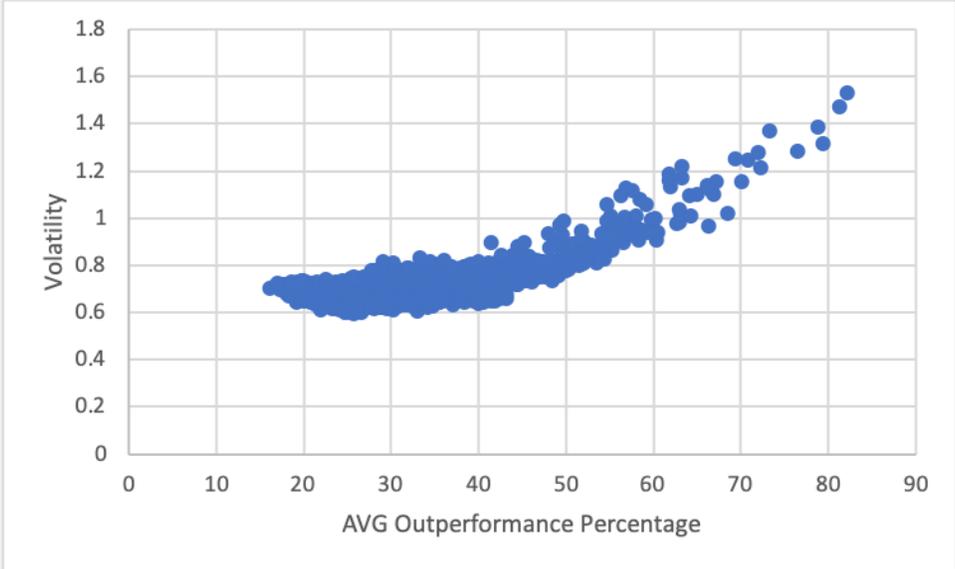

## 7. Applications and Extensions

The skills and techniques applied in this paper and many of the topics discussed aren't common knowledge to many. In public schools, there is a huge lack of financial literacy. Part of the reason that both of my papers were finance related was to help increase my peers understanding of the economy and markets that impact our everyday lives. The development and evaluation of the three models in this paper can also act as a segue into more complex financial analysis and machine learning. For others, it may remain a ground level introduction that exposes them to concepts that are helpful to know. For an investor, the evaluation of the models and how they compare against each other can inform their decision on which model to apply when investing themselves. While the models are not foolproof, they generally perform better than guessing if the stock will go up or down the following day, a 50/50 chance, or even continuously investing for a long period of time. Using models also helps with putting emotions and greed aside while trading, creating a technical and scientifically based prediction instead of a gut feeling.

This research can be further extended by attempting to increase success rates of trials and doing more extensive testing with the three models in this paper in order to create a more reliable comparison. More machine learning models can also be implemented and compared to each other. For example, a support vector classifier (SVC), which is a supervised learning classification model that categorizes data, can be used to predict stock prices and may perform better than all three of the models in this paper. Now we could implement and compare classification models with the commonly used regression models. We can also try out more technical trading strategies, like RSI (Relative Strength Index), which ranks stocks in regard to their potential value and whether or not they are "overbought".

# 9. Appendices

## A. Regression Program

```
1.  import pandas as pd
2.  import yfinance as yf
3.  import matplotlib.pyplot as plt
4.  from pandas_datareader import data as pdr
5.  import numpy as np
6.  import statistics
7.
8.  from sklearn import preprocessing as pp
9.  from sklearn.model_selection import train_test_split as tts
10. from sklearn.linear_model import LinearRegression as lr
11.
12. yf.pdr_override()
13.
14. ticker_input = input("Please enter the stock tickers you would like to invest in.\n")
15.
16. start_date = input("Here enter the date from which data will begin to be taken from (Please put it into the format of YYYY-MM-DD)\n")
17.
18. test_date = input("Here enter the date from which training and testing will end (Please put it into the format of YYYY-MM-DD)\n")
19.
20. end_date = input("Here enter the date at which we will stop taking data from (Please put it into the format of YYYY-MM-DD)\n")
21.
22. data = pdr.DataReader(ticker_input, test_date, end_date)#['Adj Close']
23. indices = data.index.tolist()
24. test_date = indices[0]
25. test_data = pdr.DataReader(ticker_input, start_date, test_date)
26.
27. def outperformance_percentage():
28.     outperformance_count = 0
29.     total_count = 0
30.     global comparison
31.
32.     for r in range (0, len(comparison.index)):
33.         if comparison[ticker_input + ' Indicative Investing'].iloc[r] > comparison[ticker_input + ' Continuous Investing'].iloc[r]:
34.             outperformance_count += 1
35.         total_count += 1
36.     outperformance_percentage = (outperformance_count / total_count) * 100
37.
38.     return outperformance_percentage
39.
40. def standard_deviation(data_list):
41.     new_list = data_list
42.     standard_deviation_sum = 0
43.     for i in range(0,len(new_list)):
44.         standard_deviation_sum += (new_list[i] - statistics.mean(new_list)) ** 2
45.     return(((standard_deviation_sum / (len(new_list) - 1)) ** (0.5)))
46.
47. #This Forecast Out will predict 1 day into the future
48. f_out = int(1)
49. test_data['Predicting'] = test_data['Adj Close'].shift(-f_out)
50.
51. X = np.array(test_data.drop(['Predicting'],1))
```

```python
52. X = pp.scale(X)
53. X_prediction = X[-f_out:]
54. X = X[:-f_out]
55.
56. Y = np.array(test_data['Predicting'])
57. Y = Y[:-f_out]
58.
59. X_train, X_test, Y_train, Y_test = tts(X, Y, test_size = 0.2)
60.
61. model = lr()
62.
63. model.fit(X_train,Y_train)
64.
65. #Prints confidence of initial model
66. conf = model.score(X_test,Y_test)
67. print("Model Confidence: ", conf)
68.
69. def investing_model(r):
70.     global test_data
71.     global test_date
72.     global data
73.     global end_date
74.     test_data = pdr.DataReader(ticker_input, start_date, test_date)
75.
76.     #This Forecast Out will predict 1 days into the future
77.     f_out = int(1)
78.     test_data['Predicting'] = test_data['Adj Close'].shift(-f_out)
79.
80.     X = np.array(test_data.drop(['Predicting'],1))
81.     X = pp.scale(X)
82.     X_prediction = X[-f_out:]
83.     X = X[:-f_out]
84.
85.     Y = np.array(test_data['Predicting'])
86.     Y = Y[:-f_out]
87.
88.     X_train, X_test, Y_train, Y_test = tts(X, Y, test_size = 0.2)
89.
90.     model = lr()
91.
92.     model.fit(X_train,Y_train)
93.
94.     prediction = model.predict(X_prediction)
95.     indices = data.index.tolist()
96.     test_date = indices[r]
97.     return(prediction - test_data.iloc[len(test_data)-1]['Adj Close'])
98.
99. last_true = data.iloc[0]['Adj Close']
100.
101.     continuous_investing = data[:]['Adj Close']
102.     indicative_investing = []
103.
104.     indicative_investing.append(last_true)
105.     for r in range (1,len(data)):
106.         x = investing_model(r)[0]
107.         if x > 0:
108.             last_true = last_true+(data.iloc[r]['Adj Close']-data.iloc[r-1]['Adj Close'])
109.         indicative_investing.append(last_true)
110.
111.     comparison = pd.DataFrame(index = data.index)
112.     comparison[ticker_input + ' Continuous Investing'] = continuous_investing
```

```
113.        comparison[ticker_input + ' Indicative Investing'] = indicative_investing
114.
115.        print("Using Moving Average Crossovers, your investment outperformed continuous investing",
116.              outperformance_percentage(),"% of the time")
117.
118.        print("The ratio of the volatility of your investment compared to continuous investing is ",
119.              standard_deviation(indicative_investing)/standard_deviation(continuous_investing))
120.
121.
122.        print(comparison)
123.        comparison.plot()
124.        plt.show()
```

### B.     KNN Program

```
1.  import pandas as pd
2.  import yfinance as yf
3.  import matplotlib.pyplot as plt
4.  from pandas_datareader import data as pdr
5.  import numpy as np
6.
7.  from sklearn import preprocessing as pp
8.  from sklearn.neighbors import KNeighborsRegressor
9.  from sklearn.model_selection import train_test_split as tts
10. from sklearn.model_selection import GridSearchCV
11. from sklearn.preprocessing import MinMaxScaler
12. scaler = MinMaxScaler(feature_range=(0,1))
13.
14. yf.pdr_override()
15.
16. ticker_input = input("Please enter the stock tickers you would like to invest in.\n")
17.
18. start_date = input("Here enter the date from which data will begin to be taken from (Please put it
    into the format of YYYY-MM-DD)\n")
19.
20. test_date = input("Here enter the date from which training and testing will end (Please put it into
     the format of YYYY-MM-DD)\n")
21.
22. end_date = input("Here enter the date at which we will stop taking data from (Please put it into th
    e format of YYYY-MM-DD)\n")
23.
24. data = pdr.DataReader(ticker_input, test_date, end_date)#['Adj Close']
25. indices = data.index.tolist()
26. test_date = indices[0]
27. test_data = pdr.DataReader(ticker_input, start_date, test_date)
28.
29. def outperformance_percentage():
30.     outperformance_count = 0
31.     total_count = 0
32.     global comparison
33.
34.     for r in range (0, len(comparison.index)):
35.         if comparison[ticker_input + ' Indicative Investing'].iloc[r] > comparison[ticker_input + '
     Continuous Investing'].iloc[r]:
36.             outperformance_count += 1
37.         total_count += 1
38.     outperformance_percentage = (outperformance_count / total_count) * 100
```

```python
39.
40.     return outperformance_percentage
41.
42. def standard_deviation(data_list):
43.     new_list = data_list
44.     standard_deviation_sum = 0
45.     for i in range(0,len(new_list)):
46.         standard_deviation_sum += (new_list[i] - statistics.mean(new_list)) ** 2
47.     return(((standard_deviation_sum / (len(new_list) - 1)) ** (0.5)))
48.
49. #This Forecast Out will predict 1 day into the future
50. f_out = int(1)
51. test_data['Predicting'] = test_data['Adj Close'].shift(-f_out)
52.
53. X = np.array(test_data.drop(['Predicting'],1))
54. X = pp.scale(X)
55. X_prediction = X[-f_out:]
56. X = X[:-f_out]
57.
58. Y = np.array(test_data['Predicting'])
59. Y = Y[:-f_out]
60.
61. X_train, X_test, Y_train, Y_test = tts(X, Y, test_size = 0.2)
62.
63. X_train_scaled = scaler.fit_transform(X_train)
64. X_test_scaled = scaler.fit_transform(X_test)
65.
66. parameters = {'n_neighbors':[2,3,4,5,6,7,8,9,10,11,12,13,14,15]}
67.
68. #define KNN algorithm and k value (k=5)
69. knn= KNeighborsRegressor()
70. model = GridSearchCV(knn, parameters, cv = 5)
71. model.fit(X_train,Y_train)
72.
73. #Prints confidence of initial model
74. conf = model.score(X_test,Y_test)
75. print("Model Confidence: ", conf)
76.
77. def investing_model(r):
78.     global test_data
79.     global test_date
80.     global data
81.     global end_date
82.     test_data = pdr.DataReader(ticker_input, start_date, test_date)
83.
84.     #This Forecast Out will predict 1 days into the future
85.     f_out = int(1)
86.     test_data['Predicting'] = test_data['Adj Close'].shift(-f_out)
87.
88.     X = np.array(test_data.drop(['Predicting'],1))
89.     X = pp.scale(X)
90.     X_prediction = X[-f_out:]
91.     X = X[:-f_out]
92.
93.     Y = np.array(test_data['Predicting'])
94.     Y = Y[:-f_out]
95.
96.     X_train, X_test, Y_train, Y_test = tts(X, Y, test_size = 0.2)
97.
98.     X_train_scaled = scaler.fit_transform(X_train)
99.     X_test_scaled = scaler.fit_transform(X_test)
```

```
100.
101.            parameters = {'n_neighbors':[2,3,4,5,6,7,8,9,10,11,12,13,14,15]}
102.
103.            #define KNN algorithm and k value (k=5)
104.            knn= KNeighborsRegressor()
105.            model = GridSearchCV(knn, parameters, cv = 5)
106.            model.fit(X_train,Y_train)
107.
108.            prediction = model.predict(X_prediction)
109.            indices = data.index.tolist()
110.            test_date = indices[r]
111.            return(prediction - test_data.iloc[len(test_data)-1]['Adj Close'])
112.
113.        last_true = data.iloc[0]['Adj Close']
114.
115.        continuous_investing = data[:]['Adj Close']
116.        indicative_investing = []
117.
118.        indicative_investing.append(last_true)
119.        for r in range (1,len(data)):
120.            x = investing_model(r)[0]
121.            if x > 0:
122.                last_true = last_true+(data.iloc[r]['Adj Close']-data.iloc[r-1]['Adj Close'])
123.            indicative_investing.append(last_true)
124.
125.        comparison = pd.DataFrame(index = data.index)
126.        comparison[ticker_input + ' Continuous Investing'] = continuous_investing
127.        comparison[ticker_input + ' Indicative Investing'] = indicative_investing
128.
129.        print("Using Moving Average Crossovers, your investment outperformed continuous investing",
130.              outperformance_percentage(),"% of the time")
131.
132.        print("The ratio of the volatility of your investment compared to continuous investing is ",
133.              standard_deviation(indicative_investing)/standard_deviation(continuous_investing))
134.
135.
136.        print(comparison)
137.        comparison.plot()
138.        plt.show()
```

### C.   Moving Average Crossover Program

```
1.  """
2.  Tests investing using Moving Average Crossover as an Indication of Entry and Exit
3.  """
4.  from pandas_datareader import data as pdr
5.  import matplotlib.pyplot as plt
6.  import pandas as pd
7.  import yfinance as yf
8.  import statistics
9.
10.
11. yf.pdr_override()
12.
13. ticker_input = input("Please enter the stock ticker you have chosen to invest in. \n")
14.
```

```python
15.    start_date= input("Here enter the date from which data will begin to be taken from (Please put it
       into format of YYYY-MM-DD)\n")
16.
17.    end_date= input("Here enter the date at which we will stop taking data from (Please put it into fo
       rmat of YYYY-MM-DD)\n")
18.
19.    ma_list = []
20.
21.    ma_value = int(input("What would you like the 1st period to be for your moving average?\n"))
22.    ma_list.append(ma_value)
23.    ma_value = int(input("What would you like the 2nd period to be for your moving average?\n"))
24.    ma_list.append(ma_value)
25.
26.
27.    #Downloads data from Yahoo using inputs given by the user.
28.    ticker_data = pdr.DataReader(ticker_input, start_date, end_date)['Adj Close']
29.
30.
31.    def simple_moving_average(length,start):
32.        s= 0
33.        for i in range (0,int(length)):
34.            #print (ticker_data[start-i])
35.            s += ticker_data[start-i]
36.        return (s/length)
37.
38.    def standard_deviation(data_list):
39.        new_list = data_list
40.        standard_deviation_sum = 0
41.        for i in range(0,len(new_list)):
42.            standard_deviation_sum += (new_list[i] - statistics.mean(new_list)) ** 2
43.        return(((standard_deviation_sum / (len(new_list) - 1)) ** (0.5)))
44.
45.    def outperformance_percentage():
46.        outperformance_count = 0
47.        total_count = 0
48.        global comparison
49.
50.        for r in range (0, len(comparison.index)):
51.            if comparison[ticker_input + ' Indicative Investing'].iloc[r] > comparison[ticker_input +
       ' Continuous Investing'].iloc[r]:
52.                outperformance_count += 1
53.            total_count += 1
54.        outperformance_percentage = (outperformance_count / total_count) * 100
55.
56.        return outperformance_percentage
57.
58.
59.    moving_averages = pd.DataFrame(index = ticker_data.index)
60.    for i in ma_list:
61.        list2 = []
62.        for v in range(0,len(ticker_data.index)):
63.            if v < i-1:
64.                list2.append(0)
65.            if v > i-1 or v == i-1:
66.                list2.append (simple_moving_average(i,v))
67.        moving_averages[ticker_input + " "+str(i) +" Day Moving Average"] = list2
68.
69.    #Remove all values that are NA and makes sure that all lists start from the same date
70.    ticker_data = ticker_data.iloc[max(ma_list)-1:]
71.    moving_averages = moving_averages.iloc[max(ma_list)-1:]
72.
```

```python
73.
74. #Calculates the difference in value between moving averages for every data-point
75. def difference_calc(x):
76.     return x[1]- x[0]
77.
78. differences=pd.DataFrame(index = ticker_data.index)
79.
80. differences = moving_averages.apply(difference_calc,axis=1)
81.
82. #Tests for sign differences between moving averages to sense crossovers
83. def test(x):
84.     if differences[x] < 0 and differences[x-1] > 0:
85.         return "sell"
86.     if differences [x] > 0 and differences[x-1] < 0:
87.         return "buy"
88.     if differences [x] > 0 and differences[x-1] > 0:
89.         return "continue_s"
90.     if differences [x] < 0 and differences[x-1] < 0:
91.         return "continue_b"
92.
93. last_true = ticker_data[0]
94.
95. continuous_investing = ticker_data[:]
96. indicative_investing = []
97.
98. if differences[0] > 0:
99.     position = "bought"
100.if differences[0] < 0:
101.    position = "sold"
102.
103.for r in range (0,len(ticker_data.index)):
104.    if position == "bought" and test(r) == "continue_b":
105.        last_true = last_true+(ticker_data[r]-ticker_data[r-1])
106.    if position == "bought" and test(r) == "sell":
107.        last_true = last_true+(ticker_data[r]-ticker_data[r-1])
108.        position == "sold"
109.    if position == "sold" and test(r) == "buy":
110.        position = "bought"
111.    indicative_investing.append(last_true)
112.
113.comparison = pd.DataFrame(index = ticker_data.index)
114.comparison[ticker_input + ' Continuous Investing'] = continuous_investing
115.comparison[ticker_input + ' Indicative Investing'] = indicative_investing
116.
117.
118.print("Using Moving Average Crossovers, your investment outperformed continuous investing",
119.      outperformance_percentage(),"% of the time")
120.
121.print("The ratio of the volatility of your investment compared to continuous investing is ",
122.      standard_deviation(indicative_investing)/standard_deviation(continuous_investing))
123.
124.moving_averages.plot()
125.differences.plot()
126.comparison.plot()
127.plt.show()
```